\documentclass{article}
\usepackage[utf8]{inputenc}

\usepackage{geometry}      
\usepackage{graphicx}
\usepackage{amssymb}
\usepackage{times}
\usepackage{color}
\usepackage{amsmath,amssymb,amsthm,graphicx,color,bm,soul}
\usepackage{enumerate}
\usepackage[normalem]{ulem}
 \newcommand{\beq}{\begin{equation}}
  \newcommand{\eeq}{\end{equation}}
  \newcommand{\beql}[1]{\begin{equation}\label{eq:#1}}

  \newcommand{\rh}{\rho}

  \newcommand{\Tr}{\mbox{\rm Tr}}
  \newcommand{\bx}{\mathbf{x}}

  \newcommand{\bS}{\mathbf{S}}
  \newcommand{\cH}{{\mathcal H}}
  \newcommand{\cI}{{\mathcal I}}
 \newcommand{\cK}{{\mathcal K}}  
\usepackage{enumerate}
\newcommand{\benum}{\begin{enumerate}[{\rm (i)}]\itemsep=0in}
\newcommand{\eenum}{\end{enumerate}}

\newcommand{\deq}[1]{\begin{align}#1\end{align}}

\newcommand{\cut}[1]{}
\newcommand{\ignore}[1]{}
\renewcommand{\star}{\ast}

\title{Quantum-like modeling in biology with open quantum systems and instruments} 
\author{Irina Basieva and Andrei Khrennikov\\
Linnaeus University, International Center for Mathematical Modeling\\  in Physics and Cognitive Sciences
 V\"axj\"o, SE-351 95, Sweden\\
Masanao Ozawa\\
College of Engineering, Chubu University, 1200 Matsumoto-cho, Kasugai 487-8501, Japan\\
Graduate School of Informatics, Nagoya University, Chikusa-ku, Nagoya 464-8601, Japan}

\date{}                     
\begin{document}
\maketitle
\abstract{We present the novel approach to mathematical modeling of information processes in biosystems. It explores the mathematical formalism and methodology of quantum theory, especially quantum measurement theory. This approach is known as {\it quantum-like} and it should be distinguished from study of genuine quantum physical processes in biosystems
(quantum biophysics, quantum cognition). It is based on quantum information representation of biosystem's state and modeling its dynamics in the framework of theory of open quantum systems. This paper starts with the non-physicist friendly presentation of quantum measurement theory, from the original von Neumann formulation to modern theory of quantum instruments. Then, latter is applied to model 
combinations of cognitive effects and gene regulation of glucose/lactose metabolism in  Escherichia  coli bacterium. The most general 
construction of quantum instruments is based on the scheme of indirect measurement, in that measurement apparatus plays the role of the environment for a biosystem. The biological essence of this scheme is illustrated by quantum formalization of Helmholtz sensation-perception theory. Then we move to open systems dynamics and consider quantum master equation, with concentrating on quantum Markov processes. In this framework, we model functioning of biological functions such as psychological functions and epigenetic mutation.} 

{\bf keywords:} mathematical formalism of quantum mechanics; open quantum systems; quantum instruments; quantum Markov dynamics;  
gene regulation; psychological effects; cognition; epigenetic mutation; biological functions
 
\section{Introduction}

The standard mathematical methods were originally developed to serve classical physics. The real analysis served as the mathematical basis of Newtonian mechanics \cite{N} (and later Hamiltonian formalism); classical statistical mechanics stimulated the measure-theoretic approach to probability theory, formalized in Kolmogorov's axiomatics \cite{K}. However, behavior of biological systems differ essentially from behavior of mechanical systems, say rigid bodies, gas molecules, or fluids. Therefore, although the ``classical mathematics'' still plays the crucial role in biological modeling, it seems  that it cannot fully describe the rich complexity of biosystems and peculiarities of their behavior -- as compared with mechanical systems. New mathematical methods for modeling biosystems are on demand.\footnote{We recall that Wigner emphasized an enormous  effectiveness of mathematics in physics \cite{Wigner}. 
But, famous Soviet mathematician Gelfand contrasted the Wigner's thesis by pointing to  
``surprising ineffectiveness of mathematics in biology'' (this remark was mentioned by another famous Soviet 
mathematician Arnold \cite{Arnold} with reference to Gelfand).}\footnote{ In particular, this special issue contains the article 
on the use of  $p$-adic numbers and analysis in mathematical biology \cite{padic}. This is non-straightforward generalization of the standard approach based on the use of real numbers and analysis.}   

In this paper, we  present the applications of the mathematical formalism of quantum mechanics and its methodology  to modeling biosystems' behavior.\footnote{We are mainly interested in quantum measurement theory (initiated in Von Neumann's book \cite{VN})  in relation with theory of quantum instruments \cite{DL70}-\cite{OO} and theory of open quantum systems \cite{OQS}.} The recent years were characterized by explosion of interest to applications of 
quantum theory outside of physics, especially in cognitive psychology, decision making, information processing in the brain, molecular biology, genetics and epigenetics, and evolution theory.\footnote{See \cite{KHC1}-\cite{KHC3} for a few pioneer papers, 
\cite{QL0} -\cite{BAF} for monographs, and \cite{HavenB}-\cite{BAX} for some representative papers.}.  We call the corresponding models {\it quantum-like}. They are not directed to micro-level modeling  of real quantum physical processes in biosystems, say in cells or brains  (cf. with 
with biological applications of genuine quantum physical theory 
\cite{P}-\cite{BER2}). Quantum-like modeling works from the viewpoint to quantum theory as a measurement theory. This is the original Bohr's viewpoint   that led to {\it the Copenhagen interpretation of quantum mechanics} (see Plotnitsky \cite{PL} for 
detailed and clear presentation of Bohr's views). One of the main bio-specialties is consideration of {\it self-measurements that  biosystems perform on themselves.} In our modeling, the ability  to perform  self-measurements is considered as the basic feature of 
biological functions (see section \ref{BF} and paper \cite{KHBR1}). 

{\it Quantum-like models} \cite{KHC3} reflect the features of biological processes that naturally match the quantum formalism.
In such modeling, it is useful to explore {\it quantum information theory,} which can be applied not just to the micro-world of quantum systems. Generally, systems processing information in the quantum-like manner need not be quantum physical systems; in particular, they can be macroscopic biosystems. Surprisingly, 
the  same mathematical theory can be applied at all biological scales: from proteins, cells and brains to  humans and ecosystems; we 
can speak about {\it quantum information biology} \cite{QIB}.

\begin{figure}[ht]
\begin{center}
\includegraphics[width=1\textwidth]{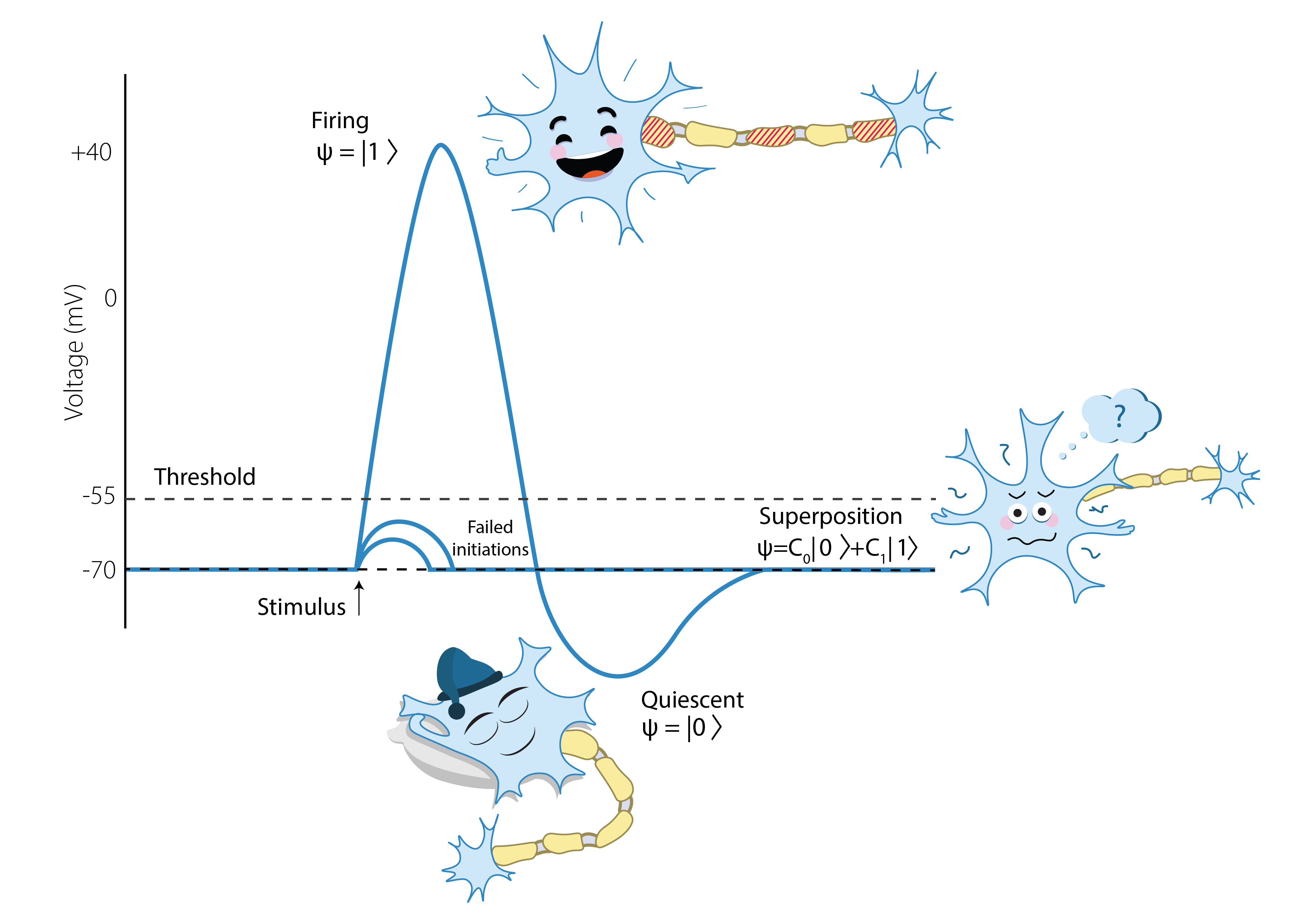}\hspace{8mm} 
\end{center}
\caption{Illustration for quantum-like representation of uncertainty generated by neuron's action potential (originally published in \cite{KHBR1}).}
\label{NEURON}
\end{figure}

In quantum-like modeling, quantum theory is considered as calculus for prediction and transformation of probabilities. Quantum probability (QP) calculus (section \ref{CPREF})  differs essentially from classical probability (CP) calculus based on  Kolmogorov's axiomatics \cite{K}. In CP, states of random systems are represented by probability measures and observables by random variables; 
in QP, states of random systems are represented by normalized vectors in a complex Hilbert space (pure states) or generally 
by density operators (mixed  states).\footnote{CP-calculus is closed calculus of probability measures. In QP-calculus \cite{VN,PR77}, probabilities are not the primary objects. They are generated from quantum states  with the aid of {\it the Born's rule.} The basic operations of QP-calculus are presented in terms of states, not probabilities. For example, the probability update 
cannot be performed straightforwardly and solely with probabilities, as done in CP - with the aid of the 
Bayes formula.} Superpositions represented by pure states are used to model uncertainty which is yet unresolved by a measurement. The use of superpositions in biology is illustrated by Fig. \ref{NEURON}  (see section \ref{ECH} and paper \cite{KHBR1} 
for the corresponding model). The QP-update resulting from an observation is based on the projection postulate or more general transformations  of quantum states - in the framework of theory of quantum instruments  \cite{DL70}-\cite{OO} (section \ref{QI}).

We stress that quantum-like modeling  elevates the role of convenience and simplicity of quantum representation of states and observables. (We pragmatically ignore the problem of interrelation of CP and QP.) In particular, the quantum state space has the linear structure and linear models are  simpler. Transition from classical 
nonlinear dynamics of electrochemical processes in  biosystems to quantum linear dynamics essentially speeds up the state-evolution (section \ref{LIN}). However, in this framework ``state'' is the quantum information state of a biosystem used for processing of special 
quantum uncertainty (section \ref{BF}).

In textbooks on  quantum mechanics, it is commonly poited out that the main distinguishing feature of quantum theory is the presence of {\it incompatible observables.} We recall that two observables, 
$A$ and $B$ are incompatible if it is impossible to assign values to them jointly. In the probabilistic model, this leads to impossibility to determine their joint probability distribution (JPD). The basic examples of incompatible observables are position and momentum of a quantum system, or spin (or polarization) projections onto different axes. In the mathematical formalism, incompatibility 
is described as noncommutativity of Hermitian operators $\hat A$ and $\hat B$  representing observables, i.e.,  $[\hat A, \hat B] \not=0.$ Here we refer to the original and still basic and widely used model of quantum observables, von Neumann \cite{VN} (section \ref{I1}). 

Incompatibility-noncommutativity is widely used in quantum physics and the basic physical observables, as say position and momentum, spin and polarization projections, are traditionally represented in this paradigm, by Hermitian operators. Still, it may be not general enough for our purpose - to quantum-like modeling in biology, not any kind of non-classical bio-statistics can be easily delegated to 
 von Neumann model of observations. For example, even very basic cognitive effects cannot be described in a way consistent with the standard observation model \cite{PLOS,FOUND}. 

We shall explore more general theory of observations based on {\it quantum instruments} \cite{DL70}-\cite{OO}  and find useful tools 
for applications to modeling of cognitive effects \cite{ENTROPY,ARXIV}. We shall discuss this question in section \ref{QI} and illustrate it with examples from  cognition and molecular biology  in sections \ref{EFF}, \ref{GEN}. In the framework of the quantum instrument theory, the crucial point is not commutativity vs. noncommutativity of operators symbolically representing observables, but the mathematical form of state's transformation resulting from the back action of (self-)observation.
In the standard approach, this transformation is given by an orthogonal projection on the subspace of eigenvectors corresponding to 
observation's output. This is {\it the projection postulate.} In quantum instrument theory, state transformations are more general.  

Calculus of quantum instruments is closely coupled with {\it theory of open quantum systems} \cite{OQS}, quantum systems interacting with environments. We remark that in some situations, quantum physical systems can be considered as (at least approximately) isolated. However, biosystems are fundamentally open. As was stressed by Schr\"odinger \cite{SCH}, a completely isolated biosystem is dead. The latter explains why the theory of open quantum systems and, in particular, the quantum instruments calculus play the basic role in applications to biology, as the mathematical apparatus of quantum information biology \cite{QIB}.
   
Within theory of open quantum systems, we model epigenetic evolution \cite{epigenetic,QLX} (sections \ref{EPGE}, \ref{ENT1}) and performance of  psychological (cognitive) functions realized by the brain \cite{IJTB,QLX,KHBR1} (sections \ref{ECH}, \ref{ENT2}). 

For mathematically sufficiently well educated biologists, but without knowledge in physics, we can recommend book \cite{PR77} 
combining the presentations of CP and QP with a brief introduction to the quantum formalism, including the theory of quantum  instruments and conditional probabilities.

\section{Classical versus quantum probability}
\label{CPREF}

CP was mathematically formalized by Kolmogorov (1933) \cite{K}.\footnote{The {\it Kolmogorov probability space} \cite{K} is any triple
$(\Omega, {\cal F}, {\bf P}),$
where $\Omega$ is a set of any origin and ${\cal F}$ is a
$\sigma$-algebra of its subsets, ${\bf P}$ is a probability measure on ${\cal F}.$
The set $\Omega$  represents random parameters of the model.
Kolmogorov called elements of $\Omega$ {\it elementary events}. Sets of elementary events  belonging to ${\cal F}$
are regarded as {\it events}.} 
 This is the
calculus of probability measures, where a non-negative weight $p(A)$ is
assigned to any event $A.$ The main property of CP is its additivity: if two events $O_1, O_2$ are
disjoint, then the probability of disjunction of these events equals to the
sum of probabilities: 
\begin{equation*}
P(O_1 \vee O_2) = P(O_1) + P(O_2). 
\end{equation*}

QP is the calculus of complex amplitudes or in the abstract formalism complex vectors. Thus, instead of operations on
probability measures one operates with vectors. We can say that QP is a 
\textit{vector model of probabilistic reasoning.} Each complex amplitude $\psi$ gives the probability by the Born's rule: \textit{Probability is
obtained as the square of the absolute value of the complex amplitude.} 
\begin{equation*}
p= \vert \psi \vert^2
\end{equation*}
(for the Hilbert space formalization, see section \ref{I1}, formula (\ref{BRRR1})).
By operating with complex probability amplitudes, instead of the direct
operation with probabilities, one can violate the basic laws of CP.

In CP, the \textit{formula of total probability} (FTP)   is derived by using 
additivity of probability and the Bayes formula, the definition of
conditional probability,
$P(O_2|O_1)=\frac{P(O_2\cap O_1)}{P(O_1)},\;P(O_1)>0.$
Consider the pair, $A$ and $B,$ of discrete  classical random variables. Then 
$$
P(B=\beta )=\sum_{\alpha }P(A=\alpha )P(B=\beta |A=\alpha ).  \label{QS7A}
$$
Thus, in CP the $B$-probability distribution can be calculated from the $A$-probability and the conditional probabilities 
$P(B=\beta|A=\alpha ).$

In QP,  classical FTP is perturbed by the interference term \cite{QL1}; 
for dichotomous quantum observables $A$ and $B$ of the von Neumann-type, i.e., given by Hermitian operators 
$\hat A$ and $\hat B,$ the quantum version of FTP has the form:
\begin{equation}
P(B=\beta )=\sum_{\alpha }P(A=\alpha )P(B=\beta |a=\alpha ) 
\end{equation}%
\begin{equation}
+2\sum_{\alpha _{1}<\alpha _{2}}\cos \theta _{\alpha _{1}\alpha _{2}}\sqrt{%
P(A=\alpha_{1})P(B=\beta |A=\alpha_{1})P(A=\alpha_{2})P(B=\beta |a=\alpha_{2})} \label{QS7AA}
\end{equation}
If the interference term\footnote{We recall that
interference is the basic feature of waves, so often one speaks about probability waves. But, one has to be careful with using 
the wave-terminology and not assign the direct physical (or biological) meaning to probability waves.} is positive, then the QP-calculus would generate a
probability that is larger than its CP-counterpart given by the classical
FTP (\ref{QS7A}). In particular, this probability amplification is the basis
of the quantum computing supremacy.

There is a plenty of statistical data from cognitive psychology, decision making, molecular biology, genetics and epigenetics demonstrating that biosystems, from proteins and cells \cite{QLX} to humans \cite{QL1,QL2} use this amplification and operate with non-CP updates. We continue our presentation with such examples.

\section{Quantum instruments} 
\label{QI}

\subsection{A few words about the quantum formalism}

Denote by ${\cal H}$ a complex Hilbert space. For simplicity, we assume that it is finite dimensional.  
Pure states of a system $S$ are given by normalized vectors of ${\cal H}$ and mixed states by density operators (positive semi-definite operators with unit trace).  The space of  density operators is denoted by $\bS(\cH).$ The space of all linear operators in ${\cal H}$ is denoted by the symbol ${\cal L}({\cal H}).$ In turn, this is a linear space.Moreover,  ${\cal L}(\cH)$ is the
complex Hilbert space with the scalar product, $\langle A\vert B\rangle= \rm{Tr} A^\star B.$  We consider linear operators acting in
 ${\cal L}(\cH).$ They are called {\it superoperators.}

The dynamics of the pure state of an isolated quantum system is described by {\it the Schr\"odinger equation:} 
\begin{equation}
\label{SCHE}
i\frac{d}{dt} \psi(t)= \hat H \psi(t)(t), \; \psi(0)= \psi_0,
\end{equation} 
where $\hat H$ is system's Hamiltonian. 
This equation implies that the pure state $\psi(t)$ evolves unitarily: $\psi(t)= \hat U(t)\psi_0,$ where $\hat U(t)= e^{-i t \hat H}$ is one parametric group of unitary operators, $\hat U(t): {\cal H} \to {\cal H}.$  In quantum physics, Hamiltonian $\hat H$ is associated with the energy-observable. The same interpretation is used in quantum biophysics \cite{QBIOP}. However, in our quantum-like modeling describing information processing in biosystems,  the operator $\hat H$ has no direct coupling  with physical energy. This is the evolution-generator describing information interactions. 

Schr\"odinger's dynamics for a pure state implies  that the dynamics of a mixed state (represented by a density operator) is described by  the {\it von Neumann equation}:
\begin{equation}
\label{VNE}
 \frac{d\hat \rho}{d t}(t)= - i [\hat  H,  \hat \rho(t)],\; \hat \rho(0)= \hat \rho_0.
\end{equation}
 
\subsection{Von Neumann formalism for quantum observables }
\label{I1}

In  the original quantum formalism \cite{VN},  physical observable $A$ is represented 
by a Hermitian operator $\hat A.$  We consider only operators with discrete spectra:
$
\hat A= \sum_x x\; \hat E^A(x),
$
where $\hat E^A(x)$ is the projector onto the subspace of $H$ corresponding to the eigenvalue $x.$ 
Suppose that system's state is mathematically represented by a density operator $\rho.$ 
Then the probability to get  the answer $x$ is given by the Born rule
\begin{equation}
\label{BRRR}
\Pr\{A =x\|\rho\} = \Tr[\hat E^A(x) \rh]= \Tr[\hat E^A(x)  \rh \hat E^A(x)]
\end{equation}
and according to the projection postulate 
the post-measurement state is obtained via the state-transformation:
\begin{equation}
\label{projection-postulate}
\rho \to \rho_x= \frac{\hat E^A(x) \rho \hat E^A(x)}{Tr \hat E^A(x) \rho \hat E^A(x)}.
\end{equation}
For reader's convenience, we present these formulas for a pure initial state $\psi \in {\cal H}.$ The Born's rule has the form:
\begin{equation}
\label{BRRR1}
\Pr\{A =x\|\rho\} = \Vert  \hat E^A(x) \psi\Vert^2= \langle \psi\vert  \hat E^A(x) \psi \rangle .
\end{equation}
The state transformation is given by the projection postulate:  
\begin{equation}
\label{BRRR2}
\psi \to \psi_x = \hat E^A(x) \psi/ \Vert  \hat E^A(x) \psi\Vert.
\end{equation}
Here the observable-operator $\hat A$ (its spectral decomposition)  uniquely determines 
the feedback state transformations $\cI_A(x)$ for outcomes $x$
\begin{equation}
\label{BRRR3}
\rho \to \cI_A(x)\rho= \hat E^A(x)\rho \hat E^A(x).
\end{equation}
The map  $x \to \cI_A(x)$ given by  (\ref{BRRR3}) is the simplest (but very important) example of quantum instrument.    

\subsection{Non-projective state update: atomic instruments}
\label{I2}

In general, the statistical properties of any measurement are characterized by 
\begin{itemize}
\item[(i)] the output probability 
distribution $\Pr\{\bx=x\| \rho\}$, the probability distribution of the output $\bx$ of the measurement
in the input state $\rho$;
\item[(ii)] the quantum state reduction $\rho\mapsto\rho_{\{\bx=x\}}$, the state change from 
the input state $\rho$ to the output state $\rho_{\{\bx=x\}}$ conditional upon the outcome 
$\bx=x$ of the measurement.
\end{itemize}
In von Neumann's formulation, the statistical properties of any measurement of an observable $A$
is uniquely determined by Born's rule (\ref{BRRR}) and the projection postulate (\ref{projection-postulate}),
and they are represented by the map (\ref{BRRR3}), an instrument of von Neumann type.
However, von Neumann's formulation does not reflect the fact that the same observable $A$ represented by
the Hermitian operator $\hat{A}$ in $\cH$ 
can be measured in many ways.\footnote{Say $\hat A= \hat H$ is the operator representing the energy-observable. This is just a theoretical entity encoding 
energy. Energy can be measured in many ways within very different measurement schemes.}
Formally, such measurement-schemes are represented by quantum instruments.  

\newcommand{\D}{\hat{D}}

Now, we consider the simplest quantum instruments of non von Neumann type,  known as 
{\it atomic instruments.}  
We start with recollection of the notion of POVM (probability operator valued measure); 
we restrict considerations to POVMs with a discrete domain of definition $X=\{x_1,...,x_N,...\}$. 
POVM is a map $x \to \D(x)$ such that  for each $x \in X,$  
$\D(x)$ is a positive contractive  Hermitian operator (called effect) (i.e., $\D(x)^\star= \D(x)$,
 $0 \leq \langle \psi\vert \D(x) \psi\rangle \leq 1$ for any $\psi \in {\cal H}$), and  the normalization condition 
$$
\sum_x \D(x)=I
$$ 
holds, where $I$ is the unit operator.
It is assumed that for any measurement,
the output probability distribution $\Pr\{\bx=x\| \rho\}$ is given by
\begin{equation}
\Pr\{\bx=x\| \rho\}=\Tr[\D(x)\rho],
\end{equation}
where $\{\D(x)\}$ is a POVM.
For atomic instruments,  it is assumed that effects are represented concretely in the form 
\begin{equation}
\label{BRULEbt} \D(x)= \hat  V(x)^\star \hat  V(x),
\end{equation}
where $V(x)$ is a linear operator in $H.$ Hence, the normalization condition has  the form $\sum_x V(x)^\star V(x) = I.$\footnote{
We remark that any \cut{orthonormal} projector $\hat E$ is Hermitian and idempotent, i.e., $\hat E^\star = \hat E$ and 
$\hat E^2=\hat E.$ Thus, any projector $\hat E^A(x)$ can be written as (\ref{BRULEbt}):  $\hat E^A(x)= \hat E^A(x)^{*}  \hat E^A(x).$
The map $x \to \hat E^A(x)$ is a special sort of POVM, the projector valued measure - PVM, the quantum instrument of the von Neumann type.} 
The Born rule can be written similarly to  (\ref{BRRR}):   
\begin{equation}
\label{BRULEb}
\Pr\{\bx =x\|\rho\} = \Tr[V(x)\rh V^\star(x)]
\end{equation}
It is assumed that the post-measurement state transformation is based on the map: 
\begin{equation}
\label{TRAr}
\rho \to \cI_A(x)\rho= V(x)\rh V^\star(x), 
\end{equation}
so the quantum state reduction is given by
\begin{equation}
\label{TRA4}
\rho \to \rho_{\{\bx=x\}}= \frac{\cI_A(x)\rho}{\Tr[ \cI_A(x)\rho]}.
\end{equation}
The map  $x \to \cI_A(x)$ given by  (\ref{TRAr}) is an atomic quantum instrument.
We remark that the Born rule (\ref{BRULEb}) can be written in the form 
 \begin{equation}
\label{BRULEy}
\Pr\{\bx =x\|\rho\} = \Tr\; [\cI_A(x) \rho].
\end{equation}

Let $\hat A$ be a Hermitian operator in $\cH.$  
Consider a POVM $\hat D= (\hat D^A(x))$ with the domain of definition given by the spectrum of $\hat A.$ This POVM represents a measurement of observable $A$ if Born's rule holds:   
\deq{\label{Born's-rule}
\Pr\{A =x\|\rho\} = \Tr[\hat D^A(x)\rh] =\Tr[\hat E^A(x)\rh].
}
Thus, in principle, probabilities of outcomes are still encoded in the spectral decomposition of operator $\hat A$ or in other words operators   $\hat D^A(x)$ should be selected in such a way that they generate the probabilities corresponding to the spectral decomposition of the symbolic representation $\hat A$ of observables $A$, i.e., $\hat D^A(x)$ is uniquely determined 
by $\hat{A}$ as
$\hat D^A(x)=\hat E^A(x)$. 
 We can say that this operator carries only information about the probabilities of outcomes, in contrast 
to the von Neumann scheme, operator $\hat A$ does not encode the rule of the state update.  
For an atomic instrument, measurements of the observable $A$ has the unique output probability distribution
by the Born's rule (\ref{Born's-rule}), but has many different quantum state reductions depending of 
the decomposition of the effect $\D(x)=\hat E^A(x)=V(x)^*V(x)$ in such a way that 
\begin{equation}
\rho\to\rho_{\{A=x\}}=\frac{V(x)\rho V(x)^*}{\Tr[V(x)\rho V(x)^*]}.
\end{equation}

\subsection{General theory (Davies--Lewis-Ozawa)} 
\label{I3}

Finally, we formulate the general notion of quantum instrument.  A superoperator acting in ${\cal L}({\cal H})$ is called positive if it maps the set of  positive semi-definite operators into itself. We remark
that, for each $x,$  $\cI_A(x)$ given by (\ref{TRAr}) can be considered as linear positive map. 

Generally any map $x \to \cI_A(x),$ where for each $x,$ the map  $ \cI_A(x)$ is a positive superoperator 
is called {\it Davies--Lewis} \cite{DL70} quantum instrument. Here index $A$ denotes the observable 
coupled to this instrument. The probabilities of $A$-outcomes are given by 
Born's rule in form (\ref{BRULEy}) and the state-update by  transformation (\ref{TRA4}).    However, Yuen \cite{Yue87} pointed out that the class of Davies--Lewis instruments is too general to exclude physically non-realizable instruments.  Ozawa \cite{Oza84} 
introduced the important additional condition to ensure that every quantum instrument
is physically realizable. This is the condition of complete positivity. A superoperator is called {\it completely positive} if its natural extension  ${\cal T} \otimes I$ to the tensor product ${\cal L}(\cH)\otimes {\cal L}(\cH)={\cal L}(\cH\otimes\cH)$ is again a positive superoperator
on ${\cal L}(\cH)\otimes {\cal L}(\cH).$  A map $x \to \cI_A(x),$ where for each $x,$ the map  $\cI_A(x)$ is a completely positive superoperator is called {\it Davies-Lewis-Ozawa} \cite{DL70,Oza84} quantum instrument or simply quantum instrument. 
As we shall see in section \ref{QI2},  complete positivity is a sufficient condition for an instrument to be physically realizable.
On the other hand,  necessity is derived as follows \cite{O4}.
Every observable $A$ of a system $S$ is identified with the observable $A\otimes I$ of a
system $S+S'$ with any system $S'$ external to $S.$
\footnotetext{For example, the color $A$ of my eyes is not only the property
of my eyes $S$ but also the property $A\otimes I$ of myself $S+S'$.}
Then, 
every physically realizable instrument ${\cal I}_A$ measuring $A$ should be identified
with the instrument ${\cal I}_{A\otimes I}$ measuring $A\otimes I$
such that ${\cal I}_{A\otimes I}(x)={\cal I}_A(x)\otimes I$.  
This implies that ${\cal I}_A(x)\otimes I$ is agin a positive superoperator,
so that ${\cal I}_A(x)$ is completely positive.  Similarly, any physically realizable instrument
$\cI_A(x)$ measuring system $S$ should have its extended instrument $\cI_A(x)\otimes I$
measuring system $S+S'$ for any external system $S'$.  This is fulfilled only
if $\cI_A(x)$ is completely positive.
Thus, complete positivity is a necessary condition for  ${\cal I}_A$ to describe a physically 
realizable instrument.

\section{Quantum instruments from the scheme of indirect measurements}
\label{QI2}

The basic model for construction of quantum instruments is based on  the scheme of indirect measurements. 
This scheme formalizes the following situation: measurement's outputs are generated via interaction of a system $S$ 
with a measurement apparatus $M.$ This apparatus consists of a complex physical device interacting with $S$ and a pointer 
that shows the result of measurement, say spin up or spin down. An observer can see only outputs of the pointer and he 
associates these outputs with the values of the observable $A$ for the system $S.$ S Thus, the indirect measurement scheme involves:
\begin{itemize}
\item the states of the systems $S$ and the apparatus $M;$    
\item the operator $U$ representing the interaction-dynamics for the system $S+M;$ 
\item the meter observable $M_A$  giving outputs of the pointer of the apparatus $M.$ 
\end{itemize}

An {\em indirect measurement model}, introduced in \cite{Oza84} as a ``(general) measuring process'',
is a quadruple 
$$
(\cK, \sigma, U, M_A)
$$ 
consisting of a Hilbert space $\cK,$   a density operator $\sigma \in {\bf S}(\cK),$ a unitary operator $U$ 
on the tensor product of the state spaces of $S$ and $M,$ $U: \cH\otimes\cK \to \cH\otimes\cK,$ and a Hermitian operator $M_A$ on $\cK$.  
By this measurement model,  the Hilbert space $\cK$ describes the states of the apparatus  $M$, the unitary operator  $U$ describes the time-evolution of the composite system $S+ M$, the density operator $\sigma$ describes the initial state of the apparatus $M$, and the Hermitian operator $M_A$ describes the meter observable of the apparatus $M.$ Then, the output probability distribution $\Pr\{A =x\|\rho\}$ in the system state $\rho \in \bf{S}(\cH)$
is given by
\beq
\label{MA1}
\Pr\{A =x\|\rho\}=\Tr[(I\otimes E^{M_A}(x))U(\rho\otimes\sigma)U^{\star}],
\eeq 
where $E^{M_A}(x)$ is  the spectral projection of $M_A$ for the eigenvalue $x.$

The change of the state $\rho$ of the system $S$  caused by the measurement for the outcome $A =x$ 
is represented with the aid of the map $\cI_A(x)$ in the space of density operators defined as 
\beq
\label{MA1A}
\cI_A(x)\rho =\Tr_{\cK}[(I\otimes E^{M_A}(x))U(\rho\otimes\sigma)U^{\star}],
\eeq 
where $\Tr_{\cK}$ is the partial trace over $\cK.$ 
Then,  the map  $x \mapsto \cI_A(x)$ turn out to be a quantum instrument. 
Thus, the statistical properties of the measurement realized by any indirect measurement 
model $(\cK,\sigma,U,M_A)$ is described by a quantum measurement.
We remark that conversely any quantum instrument can be represented via the indirect 
measurement model \cite{Oza84}.   Thus, quantum instruments mathematically 
characterize the statistical properties of all the physically realizable quantum 
measurements.

\section{Modeling of the process of sensation-perception within indirect measurement scheme}

Foundations of theory of {\it unconscious inference} for the formation of visual impressions were set in 19th century by H. von Helmholtz.  Although von Helmholtz studied mainly  visual sensation-perception, he also applied his theory for other senses up to culmination in theory of social unconscious inference. By von Helmholtz here are two stages of the cognitive process, and they discriminate between \textit{sensation} and \textit{perception} as follows:
\begin{itemize}
\item 
Sensation is a signal which the brain interprets as a sound or visual image, etc.
\item 
Perception is something to be interpreted as a preference or selective attention, etc.
\end{itemize}
In the scheme of  indirect measurement, sensations represent the states of the sensation system $S$ of human and the perception 
system plays the role of  the measurement apparatus $M.$ The unitary operator $U$ describes the process of interaction between 
the sensation and perception states. This quantum modeling of the process of sensation-perception was presented in paper \cite{FRONTIERS}
with application to bistable perception and experimental data from article \cite{Bell2}.
   
\section{Modeling of cognitive effects}
\label{EFF}     
  
In cognitive and social science, the following opinion pool is known as the basic example of the order effect. 
 This is  the Clinton-Gore opinion pool \cite{Moore}. In this experiment, American citizens  were
asked one question at a time, e.g., 
\begin{itemize}
\item $A$ = ``Is Bill Clinton honest and trustworthy?'' 
\item $B$  = ``Is Al Gore honest and trustworthy?''
\end{itemize}
Two sequential probability distributions were calculated on the basis of the experimental statistical data, $p_{AB}$ and $p_{BA}$ (first question $A$ and then question $B$ and vice verse).  

\subsection{Order effect for sequential questioning}

The statistical data from this experiment  demonstrated the {\it question order effect} QOE, dependence of sequential joint probability distribution for answers to the questions on their order: $p_{AB}\not=p_{BA}.$ We remark that in the CP-model 
these probability distributions coincide: $p_{AB}(\alpha, \beta)=P(\omega \in \Omega: A(\omega)=\alpha, B(\omega)=\beta)=
p_{BA}(\beta,\alpha),$ where $\Omega$ is a sample space and $P$ is a probability measure.    
QOE stimulates application of  the QP-calculus to cognition, see paper \cite{WB13}. The authors of this paper  
stressed that noncommutative feature of joint probabilities can be modeled by using noncommutativity of incompatible quantum observables 
$A, B$ represented by Hermitian operators  $\hat A, \hat B.$  Observable $A$ represents the Clinton-question and observable $B$ represents Gore-question.  In this model,  QOE is identical incompatibility-noncommutativity of observables:
$[\hat A, \hat B]\not=0.$

\subsection{Response replicability effect for sequential questioning}
The approach based on identification of the order effect with noncommutative representation of questions \cite{WB13} was
criticized in  paper \cite{PLOS}. To discuss this paper, we recall  the notion of {\it response replicability.}
 Suppose that a person, say John,  is asked some question $A$ and suppose that he replies, e,g, ``yes''.  If immediately after this, he is asked the same question again, then he replies ``yes''  with probability one.
We call this property   {\it $A-A$ response replicability.} In quantum physics, $A-A$ response replicability is  expressed by  {\it the projection postulate.} The Clinton-Gore opinion poll as well as typical decision making experiments satisfy $A-A$ response replicability. 
Decision making has also another feature -  {\it $A-B-A$ response replicability.}  Suppose that after answering the $A$-question with say the ``yes''-answer, John is  asked another question $B.$ He replied to it with some answer. 
And then he is asked $A$ again.  In the aforementioned social opinion pool,  John  repeats her original answer to $A,$ ``yes'' (with probability one). This behavioral phenomenon we call $A-B-A$ response replicability.  Combination of $A-A$ with $A-B-A$ and $B-A-B$ response replicability is called {\it the response replicability effect} RRE.

\subsection{``QOE+RRE'':  described by quantum instruments of non-projective type}
In paper \cite{PLOS}, it was shown that by using the von Neumann calculus it is {\it impossible to combine  RRE  with QOE.}  
To generate QOE, Hermitian operators $\hat A, \hat B$ should be noncommutative, but the latter destroys$A-B-A$ response replicability  of $A.$
This was a rather unexpected result. It made even impression that, although the basic cognitive effects can be quantum-likely 
modeled separately, their combinations cannot be described by the quantum formalism.  

However, recently it was shown  that theory of quantum instruments provides a simple solution of the combination of QOE and RRE effects, 
see \cite{ENTROPY} for construction of such instruments. These instruments are of non-projective type. Thus, the essence of QOE is not  in the structure of observables, but in the structure of the state transformation generated by measurements' feedback. QOE is not about the joint measurement and incompatibility (noncommutativity) of observables, but about sequential measurement of observables and sequential (mental-)state update. Quantum instruments which are used in \cite{ENTROPY} to combine QOE and RRE correspond to  {\cal measurement of  observables} represented by commuting operators   $\hat A, \hat B.$  Moreover, it is possible to prove that (under natural mathematical restriction) QOE and RRE can be jointly modeled only with the aid of quantum instruments for commuting observables.          

\subsection{Mental realism} 

Since very beginning of quantum mechanics, noncommutativity of operators $\hat A, \hat B$ representing observables 
$A, B$ was considered as the mathematical representation of their incompatibility. In philosophic terms, this situation is treated as impossibility of the realistic description. In cognitive science, this means that there exist mental states such that
an individual cannot assign the definite values to both observables (e.g., questions). The mathematical description  of  QOE with observables  represented by noncommutative operators (in the von Neumann's scheme) in \cite{WB13,WSSB14} made impression that this effect implies rejection of mental realism. The result of \cite{ENTROPY} demonstrates that, in spite of experimentally well documented QOE, the mental realism need not be rejected. QOE can be modeled within the realistic picture mathematically given by the joint probability distribution of observables $A$ and $B,$ but with the noncommutative action of quantum instruments updating the mental state:
\beq
\label{MA1AT}
[\cI_A(x), \cI_B(x)]= \cI_A(x) \cI_B(x) -  \cI_B(x) \cI_A(x)\not=0.
\eeq 
This is the good place to remark that if, for some state $\rho, [\cI_A(x), \cI_B(x)] \rho=0,$ then QOE disappears, even if
$[\cI_A(x), \cI_B(x)]\not=0.$ This can be considered 
as the right formulation of Wang-Bussemeyer statement on connection of QOE with noncommutativity. Instead of noncommutativity of operators $\hat A$ and $\hat B$ symbolically representing quantum obseravbles, one has to speak about noncommutativity of corresponding 
quantum instruments. 

\section{Genetics: interference in glucose/lactose metabolism}
\label{GEN}

In paper \cite{EColi}, there was developed a  quantum-like  model  describing the gene regulation of glucose/lactose metabolism in  Escherichia  coli bacterium.\footnote{This is the first application of  the operational quantum formalism  for microbiology  and  genetics.  Instead  of  trying  to  describe the physical and biological processes in a cell in the very detail, there was constructed the formal operator representation. Such representation is useful when the detailed description of processes is very complicated. In the aforementioned paper,  the statistical  data  obtained  from  experiments in genetics was analyzed from the CP-viewpoint;  the degree of E. coli’s preference within adaptive dynamics to glucose/lactose environment was  computed.} There are several types of E. coli characterized  by  the  metabolic  system.  It was  demonstrated  that  the concrete  type  of E.  coli can  be  described  by  the  well  determined   linear operators;   we   find  the invariant   operator   quantities characterizing each type. Such invariant operator quantities can be calculated from the obtained statistical data. So, the quantum-like representation was reconstructed from experimental data. 

Let us consider an  event system $\{Q_+,Q_-\}: Q_+$ means the event that E. coli activates its lactose operon, that is, the event that 
$\beta$-galactosidase is produced through the transcription of mRNA from a gene in lactose operon; $Q_-$ means the event that E. coli does not activates its lactose operon. This system of events corresponds to activation observable $Q$ that is mathematically represented
by a quantum instrument ${\cal T}_Q.$  Consider now another system of events $\{D_L,D_G\}$ where  $D_L$  means the event that an E. coli bacterium detects a lactose molecular in cell's surrounding environment, 
$D_G$ means detection of a glucose molecular. This system of events corresponds to detection observable $D$ that is represented by a quantum instrument ${\cal T}_D.$    In this model, bacterium's interaction-reaction with glucose/lactose environment is described as sequential action of two quantum  instruments, first detection and then activation. As was shown in \cite{EColi}, 
for each concrete type of E. coli bacterium, these quantum instruments can be reconstructed from the experimental data; 
in \cite{EColi},  reconstruction was performed for W3110-type of E. coli bacterium. The classical FTP with observables $A=D$ and $B=Q$ is violated, the interference term, see (\ref{QS7AA}), was calculated \cite{EColi}.  

\section{Open quantum systems: interaction of a biosystem with its environment}

As was already emphasized, any biosystem $S$ is fundamentally open. Hence, dynamics of 
its state has to be modeled via an interaction with surrounding environment ${\cal E}.$ 
The states of $S$ and ${\cal E}$ are represented in  the Hilbert spaces $\cH$ and $\cK.$
The compound system $S+ {\cal E}$ is  represented in  the tensor product 
Hilbert spaces $\cH \otimes \cK.$ This system is treated as an isolated system and in accordance with quantum theory, dynamics 
of its pure state can be described by the Schr\"odinger equation: 
\begin{equation}
\label{SCHET}
i\frac{d}{dt} \Psi(t)= \hat H \Psi(t)(t), \; \Psi(0)= \Psi_0,
\end{equation} 
where $\Psi(t)$ is the pure state of the system  $S+ {\cal E}$ and $\hat H$ is its Hamiltonian. 
This equation implies that the pure state $\psi(t)$ evolves unitarily : $\Psi(t)= \hat U(t)\Psi_0,$ where $\hat U(t)= e^{-i t \hat H}.$ Hamiltonian, the evolution-generator, describing information interactions has the form $\hat H =\hat H_S + \hat H_{{\cal E}} + \hat H_{S,{\cal E}},$ where $\hat H_S, \hat H_{{\cal E}}$ are Hamiltonians of the systems and $\hat H_{S,{\cal E}}$ is the interaction Hamiltonian.\footnote{ The unitary evolution of the state of the compound system $S+ {\cal E}$ is incorporated in the indirect measurement scheme (section \ref{QI2}).}  This equation implies that evolution of the density operator $\hat R(t)$ of the system $S+ {\cal E}$ is described by von Neumann equation: 
\begin{equation}
\label{VNEE}
 \frac{d\hat R}{d t}(t)= - i [\hat  H,  \hat R(t)],\; \hat R(0)= \hat R_0,
\end{equation}
However, the state $\hat R(t)$ is too complex for any mathematical analysis: the environment includes too many degrees of freedom.
Therefore, we are interested only the state of $S;$ its dynamics is obtained via tracing of the state of  $S+ {\cal E}$  
w.r.t. the degrees of freedom of ${\cal E}:$ 
\begin{equation}
\label{VNEE1}
 \hat \rho(t) = \rm{Tr}_{\cK} \hat R(t).
\end{equation}
Generally this equation, {\it the quantum master equation},  is mathematically very complicated. A variety of approximations is used in applications.

\subsection{Quantum Markov model: Gorini-Kossakowski-Sudarshan-Lindblad equation}
\label{QM}

The simplest approximation of quantum master equation (\ref{VNEE1})
is {\it the quantum  Markov dynamics} given by the {\it Gorini-Kossakowski-Sudarshan-Lindblad} (GKSL) equation \cite{OQS}
(in physics, it is commonly called simply the Lindblad equation; this is the simplest quantum master equation):
\begin{equation}
\label{GKSL}
 \frac{d\hat \rho}{d t}(t)= - i [\hat  H,  \hat \rho(t)] +  \hat L [\hat \rho(t)], \; \hat \rho(0)= \hat \rho_0,
\end{equation}
where Hermitian operator (Hamiltonian) $\hat  H$ describes the internal dynamics of $S$ and the superoperator $\hat L,$ acting 
in the space of density operators, describes an interaction with environment  ${\cal E}.$ This superoperator is often called 
{\it Lindbladian.} The GKSL-equation is a quantum master equation for Markovian dynamics. In this paper, we have no possibility 
to explain the notion of quantum Markovianity in more detail. Quantum master equation (\ref{VNEE1}) describes generally non-Markovean dynamics.    

\subsection{Biological functions in the quantum Markov framework}
\label{BF}

We turn to the open system dynamics with the GKSL-equation. In our modeling,  Hamiltonian  $\hat H$ and Lindbladian $\hat L$ represent
some special {\it biological function} $F$ (see \cite{KHBR1}) for details. Its functioning results from interaction of internal and external information flows. 
In sections \ref{ECH}, \ref{ENT2}, $F$ is some {\it psychological function}; in the simplest case $F$ represents a question asked to $S$ (say $S$ is a human being). In section \ref{GEN}, $F$ is  the {\it gene regulation} of glucose/lactose metabolism in  Escherichia  coli bacterium. In sections \ref{EPGE}, \ref{ENT1}, $F$ represents the process of {\it epigenetic mutation}. Symbolically biological function $F$ is represented as a quantum observable:  Hermitian operator $\hat F$ with the spectral decomposition $\hat F= \sum_x x \hat E^F (x),$
where $x$ labels outputs of $F.$ Theory of quantum Markov state-dynamics describes the process of generation of these outputs.   
 
In the mathematical model \cite{QLX,AS3,AS2017,QIB,epigenetic,IJTB,EColi}, the outputs of biological function $F$ are generated via approaching  a {\it steady  state} 
of the GKSL-dynamics:
\begin{equation}
\label{SCH2}
\lim_{t \to \infty} \hat   \rho(t)= \hat   \rho_{\rm{steady}} 
\end{equation}
such that it matches the spectral decomposition of $\hat F,$ i.e., 
\begin{equation}
\label{SCH3}
\hat   \rho_{\rm{steady}} = \sum_x p_x \hat E^F (x), \; \mbox{where}\; p_x \geq 0,   \sum_x p_x =1.
\end{equation}
This means that $\hat   \rho_{\rm{steady}}$ is diagonal in an orthonormal basis consisting of eigenvectors of $\hat F.$ 
This state, or more precisely, this decomposition of density operator $\hat   \rho_{\rm{steady}}$, is the 
classical statistical mixture of the basic information states determining this biological function.
The probabilities in state's decomposition (\ref{SCH3}) are interpreted statistically. 

Consider a large ensemble of biosystems with the state $\hat   \rho_0$ interacting with environment ${\cal E}.$ (We recall that
mathematically the interaction is encoded in the Lindbladian $\hat L.)$ Resulting from this interaction, biological function $F$ 
produces output $x$ with  probability $p_x.$ We remark that in the operator terms the probability is expressed as
$p_x = \rm{Tr} \hat   \rho_{\rm{steady}} \hat E^F (x).$  This interpretation can be applied even to a single biosystem that meets the same environment many times.

It should be noted that   limiting  state $\hat \rho_{\rm{steady}}$   expresses  the  stability  with  respect  to the  influence  of  concrete  environment ${\cal E}.$ Of course, in the real world the limit-state would be never approached. The mathematical formula 
(\ref{SCH2}) describes the process of stabilization, damping of fluctuations. But, they would be never disappear completely with time.  

We note that a steady state satisfies the stationary   GKSL-equation:
\begin{equation}
\label{GKSLS}
 i [\hat  H,  \hat \rho_{\rm{steady}}] =  \hat L [\hat \rho_{\rm{steady}}].
\end{equation}
It is also important to point that generally a steady state of  the quantum master equation is not unique, it depends on the class of initial conditions.   

\subsection{Operation of biological functions through decoherence}
\label{COG}

To make the previous considerations concrete, let us consider a pure quantum state as the initial state. Suppose that a biological function $F$ is dichotomous, $F=0,1,$ and it is symbolically represented by the Hermitian operator that is diagonal in 
orthonormal basis $\vert 0 \rangle, \vert 1 \rangle.$ (We consider the two dimensional state space - the qubit space.) 
Let the initial state has the form  of superposition 
\begin{equation}
\label{SP}
\vert \psi \rangle=  c_0 \vert 0 \rangle + c_1 \vert 1 \rangle,  
\end{equation}
where $c_j \in \mathbf{C}, \vert c_0\vert^2 + \vert c_1\vert^2 =1.$ 
 The quantum master dynamics is not a pure state dynamics:  sooner or later (in fact, very soon), this superposition representing a pure state  will be transferred into  a density matrix representing a mixed state.
Therefore, from the very beginning it is useful to represent superposition (\ref{SP}) in terms of a density matrix:
\begin{equation}
\label{DM1}
\hat \rho_0 =\begin{vmatrix}
 \vert c_0\vert^2 & c_0 \bar{c}_1\\
\bar{c}_0 c_1 & \vert c_1\vert^2 \\
\end{vmatrix}
\end{equation}
State's purity,  superposition, is characterized by the presence of nonzero  off-diagonal terms.

Superposition encodes uncertainty with respect to the concrete state basis, in our case   $\vert 0 \rangle, \vert 1 \rangle.$ 
Initially biological function $F$ was in the state of uncertainty between two choices $x=0,1.$ This is {\it genuine quantum(-like) uncertainty.} Uncertainty, about possible actions in future. For example, for psychological function (section \ref{ECH}) $F$  representing answering to some question, say ``to buy property'' ($F=1)$ and its negation ($F=0),$ a person whose state is described by superposition  
(\ref{SP}) is uncertain to act with $F=1$ or with $F=0.$ Thus, a superposition-type state describes {\it individual uncertainty,} i.e., 
uncertainty associated with the individual biosystem and not with an ensemble of biosystems; with the single act of functioning of $F$ and not with a large series of such acts.   

Resolution of uncertainty with respect to  $\hat F$-basis is characterized by washing off the off-diagonal terms in   (\ref{DM1})
The quantum dynamics (\ref{GKSL}) suppresses the off-diagonal terms and, finally, a diagonal density matrix representing a steady state of  this dynamical systems is generated:
\begin{equation}
\label{DM2}
\hat \rho_{\rm{steady}} =\begin{vmatrix}
 p_{0} & 0\\
 0 & p_{1} \\
\end{vmatrix}
\end{equation}
This is a classical statistical mixture. It describes an ensemble of biosystems; statistically they generate outputs $F=\alpha$ with probabilities $p_\alpha.$ In the same way, the statistical interpretation can be used for a 
single system that performs $F$-functioning at different instances of time (for a long time series). 

In quantum physics, the process of washing off the off-diagonal elements in a density matrix 
is known as the {\it process of decoherence.}  Thus, the described model of can be called  operation of biological function 
through decoherence.  

\subsection{Linearity of quantum representation: exponential speed up for biological functioning}
\label{LIN}

The quantum-like modeling does not claim that biosystems are fundamentally quantum. 
A more natural picture is that they are a complex classical  biophysical systems and the quantum-like
model provides the information representation of classical biophysical processes, in genes, proteins, cells, 
brains. One of the advantages of this representation is its linearity. The quantum state space is a complex Hilbert space
and dynamical equations are linear differential equations. For finite dimensional state spaces, these are just ordinary differential
equations with complex coefficients (so, the reader should not be afraid of such pathetic names as Schr\"odinger, von Neumann,
or Gorini-Kossakowski-Sudarshan-Lindblad equations). The classical biophysical dynamics beyond the quantum information representation 
is typically nonlinear and very complicated. The use of the linear space representation simplifies the processing structure. There are two viewpoints on this simplification, external and internal. The first one is simplification of mathematical modeling, i.e., simplification of study of bioprocesses (by us, external observers). The second one is more delicate and interesting.  We have already pointed to one 
important specialty of applications of the quantum theory to biology. Here, systems can perform {\it self-observations.} So, in the process of evolution say  a cell can ``learn'' via such self-observations that it is computationally profitable to use the linear 
quantum-like representation. And now, we come to the main advantage of linearity. The linear dynamics exponentially speeds up 
information processing. Solutions of  the GKSL-equation can be represented in the form
$\hat \rho(t)= e^{t \hat \Gamma} \hat \rho_0,$ where $ \hat \Gamma$ is the superoperator given by the right-hand side of the 
GKSL-equation. In the finite dimensional case, decoherence dynamics is expressed via  factors
of the form $e^{t (i a- b)},$ where $b >0.$ Such factors are exponentially decreasing. Quantum-like linear realization of biological functions is exponentially rapid comparing with nonlinear classical dynamics. 

The use of the quantum information representation means that generally large clusters of classical biophysical states are encoded 
by a few quantum states. It means huge information compressing. It also implies increasing of stability in   
state-processing. Noisy nonlinear classical dynamics is mapped to dynamics driven by linear quantum(-like) equation of say 
GKSL-type. The latter has essentially simpler structure and via selection of the operator coefficients encoding symbolically 
interaction within the  system $S$ and with its surrounding environment ${\cal E},$ $S$ can establish dynamics with 
stabilization regimes leading to steady states.

\section{Epigenetic evolution within theory of open quantum systems}
\label{EPGE}

In paper \cite{epigenetic}, a general model of the epigenetic evolution unifying neo-Darwinian with neo-Lamarckian approaches was created in the 
framework of  theory of open quantum systems.  The process of evolution  is  represented  in  the  form  of {\it adaptive  dynamics}  given  by  the  quantum(-like)  master equation describing the dynamics of  the information  state  of  epigenome  in the process of interaction with surrounding environment.  This model of the epigenetic evolution expresses the  probabilities  for  observations which can be done on epigenomes of cells;  this (quantum-like) model does not give a detailed description of cellular processes.   The   quantum operational   approach   provides   a   possibility   to describe by  one model  all  known  types  of cellular  epigenetic inheritance. 

To give some hint about the model, we consider one  gene, say $g.$ This is the system $S$ in section \ref{QM}. It interacts  with the surrounding environment ${\cal E}:$ a cell containing this gene  and other cells that send signals to this concrete cell and through it to the gene $g.$  As a consequence of this interaction  some  epigenetic mutation $\mu$ in the gene $g$ can happen. It would change the level of the $g$-expression. For the moment, we  ignore  that there are other genes.
In this oversimplified model, the mutation can be  described within the two dimensional state space, complex Hilbert space 
${\cal H}_{\rm{epi}}$ (qubit  space).  States  of $g$ without  and with mutation are represented by the  orthogonal
basis $\vert 0 \rangle, \vert 1 \rangle;$ these vectors express  possible  epigenetic  changes  of  the fixed type $\mu.$
A pure quantum information state has the form of superposition 
$\vert \psi_{\rm{epi}} \rangle=  c_0 \vert 0 \rangle + c_1 \vert 1 \rangle.$ Now, we turn to the general scheme of section \ref{BF}
with the biological function $F$ expressing $\mu$-epimutation in one fixed gene. The quantum Markov dynamics (\ref{GKSL}) resolves uncertainty encoded in superposition $\vert \psi_{\rm{epi}} \rangle$ (``modeling  epimutations as decoherence'').  The classical statistical mixture  $\rho_{\rm{steady}},$ see (\ref{DM2}), is approached.  Its  diagonal elements $p_{0}, p_{1}$ give the  probabilities  of  the events:  ``no $\mu$-epimutation''  and ``$\mu$-epimutation''.  These probabilities are interpreted statistically:   in  a large population of cells, $M$ cells, $M\gg 1,$, the number of cells with $\mu$-epimutation is $N_m \approx p_{1} M.$     This $\mu$-epimutation in a cell  population  would   stabilize completely to the steady state  only in the  infinite  time. Therefore in reality there are fluctuations (of decreasing amplitude) in  any finite interval of time. 

Finally, we point to the advantage of the quantum-like dynamics of interaction of genes with environment - dynamics' linearity implying 
exponential speed up of the process of epigenetic evolution (section \ref{LIN}).  

\section{Connecting electrochemical processes in neural networks with quantum informational 
processing}
\label{ECH}

As was emphasized in introduction, quantum-like models are formal operational models describing information processing in biosystems.
(in contrast to studies in quantum biology - the science about the genuine quantum physical processes in biosystems). Nevertheless, it is 
interesting to connect the structure quantum information processing in a biosystem with physical and chemical processes in it. This is a problem of high complexity. Papers \cite{KHBR1} present the attempt to proceed in this direction for 
the human brain - the most complicated biosystem (and at the same time the most interesting for scientists). In the framework of quantum information theory,  there was modeled information processing by brain's neural networks. The quantum information formalization of the states of neural networks is coupled with the electrochemical processes in the brain. The key-point is representation of uncertainty generated by the action potential of a neuron as quantum(-like) superposition 
of the basic mental states corresponding to a neural code, see Fig. \ref{NEURON}  for illustration. 

Consider information processing by a single neuron; this is the system $S$ (see section \ref{BF}). Its quantum information state corresponding to the neural code {\it quiescent and firing,} 0/1, can be represented in the two dimensional complex Hilbert space ${\cal H}_{\rm{neuron}}$ (qubit space).  
At a concrete instant of time neuron's state can be mathematically described by superposition of two states, 
labeled by  $\vert 0\rangle, \vert 1\rangle:$
$
\vert \psi_{\rm{neuron}} \rangle = c_0 \vert 0\rangle+c_1 \vert 1\rangle.
$    
It is assumed that these states are orthogonal and normalized, i.e., $\langle 0 \vert 1\rangle =0$ and  $\langle \alpha \vert \alpha\rangle =1, \alpha=0,1.$ The coordinates  $c_0$ and $c_1$  with respect to the quiescent-firing basis  are complex amplitudes representing potentialities for the neuron $S$  to  be quiescent or firing.  Superposition represents uncertainty in action potential, 
``to fire'' or ``not to fire''. This superposition is quantum information representation of physical, electrochemical uncertainty. 

Let $F$ be some {\it psychological (cognitive) function} realized by this neuron. (Of course, this is oversimplification, considered, e.g., in the paradigm  ``grandmother neuron''; see section \ref{ENT2} for modeling of $F$ based on a neural network). We assume that $F=0,1$ is dichotomous. Say $F$ represents some instinct, e.g., aggression: ``attack''=1, ``not attack''=0. A psychological function can represent answering to some question (or class of questions), solving problems, performing tasks. Mathematically $F$   is represented by the Hermitian operator $\hat F$ that is diagonal in the basis $\vert 0\rangle, \vert 1\rangle.$ The neuron $S$ interacts with the surrounding electrochemical environment ${\cal E}.$ This
interaction generates the evolution of neuron's state and realization of the psychological function $F.$  We model dynamics with the quantum master equation  (\ref{GKSL}). Decoherence transforms the pure state  $\vert \psi_{\rm{neuron}} \rangle$ into the classical statistical mixture (\ref{DM2}), a steady state of this dynamics. This is resolution of the original electrochemical uncertainty
in neuron's action potential.  

The diagonal elements  of $\hat \rho_{\rm{steady}}$ give the probabilities with the statistical interpretation: in a large ensemble of neurons (individually) interacting with the same environment ${\cal E},$ say $M$ neurons, 
$M>>1,$   the number of neurons which take the decision $F=1$ equals to the diagonal element $p_1.$    

We also point to the advantage of the quantum-like dynamics of the interaction of a neuron with its environment - dynamics' linearity implying exponential speed up of the process of neuron's state evolution towards a ``decision-matrix'' given by a steady state  
(section \ref{LIN}).  

\section{Compound biosystems}

\subsection{Entanglement of information states of biosystems}
\label{ENT3}

The state space ${\cal H}$ of the  biosystem $S$ consisting of the subsystems $S_j, j=1,2,..,n,$ is the tensor product of
of subsystems' state spaces ${\cal H}_j,$ so 
\begin{equation}
\label{BE0}
{\cal H}= {\cal H}_1 \otimes ... \otimes {\cal H}_n.
\end{equation}
 The easiest way to imagine
this state space is to consider its coordinate representation with respect to some basis constructed with bases in ${\cal H}_j.$
For simplicity, consider the case of qubit state spaces   ${\cal H}_j;$ let $\vert \alpha \rangle, \alpha=0,1,$ be some orthonormal basis 
in ${\cal H}_j,$ i.e., elements of this space are linear combinations of the form $\vert \psi_j\rangle = c_0 \vert 0 \rangle+
c_1 \vert 1 \rangle.$ (To be completely formal, we have to label basis vectors with the index $j,$ i.e., $\vert \alpha \rangle\equiv 
\vert \alpha \rangle_j.$ But we shall omit this it.) Then vectors $\vert \alpha_1 ... \alpha_n \rangle\equiv  \vert \alpha_1 \rangle \otimes ...\otimes \vert \alpha_n \rangle$ form the orthonormal basis in  ${\cal H},$ i.e., any state $\vert \Psi \rangle \in {\cal H}$
can be represented in the form 
\begin{equation}
\label{BE}
\vert \Psi \rangle  = \sum_{\alpha_j=0,1} C_{\alpha_1...\alpha_n} \vert \alpha_1 ... \alpha_n \rangle,
\end{equation}
and the complex coordinates $C_{\alpha_1...\alpha_n}$ are normalized: $\sum \vert C_{\alpha_1...\alpha_n}\vert^2=1.$
For example, if $n=2,$ we can consider the state
\begin{equation}
\label{BE1}
\vert \Psi \rangle  =   (\vert 00\rangle+ \vert 11\rangle)/\sqrt{2}. 
\end{equation}
This is an example of an {\it entangled state}, i.e., a state that cannot be factorized in the tensor product of the states of   
the subsystems. An example of a non-entangled state  (up to normalization) is given by
$$
\vert 00\rangle+ \vert 01\rangle + \vert 10\rangle + \vert 11\rangle= (\vert 0\rangle + \vert 1\rangle) \otimes (\vert 1\rangle + 
\vert 0\rangle).
$$
Entangled states are basic states for quantum computing that explores state's inseparability. Acting to one concrete qubit modifies 
the whole state. For a separable state, by transforming say the first qubit, we change only the state of system $S_1.$  
This possibility to change the very complex state of a compound system via change of the local state of a subsystem is considered as the root of superiority of quantum computation over classical one.  We remark that the dimension of the tensor product state space is very big,  it equals $2^n$ for $n$ qubit subsystems. In quantum physics, this possibility to manipulate with the compound state (that can have 
the big dimension) is typically associated with ``quantum nonlocality'' and {\it spooky action at a distance.} But, even in quantum physics this nonlocal interpretation is the source for permanent debates \cite{}. In particular, in the recent series of papers \cite{}  
it was shown that it is possible to proceed without referring to quantum nonlocality and that quantum mechanics can be interpreted as the local physical theory. {\it The local viewpoint on the quantum theory is more natural for biological application.}\footnote{``Local'' with respect to the physical space-time.}  For biosystems, spooky action at a distance is really mysterious; for humans, it corresponds to acceptance of parapsychological phenomena.
 
How  can one explain generation of  state-transformation of the compound system $S$ by ``local transformation'' of say the state 
of its subsystem $S_1?$ Here the key-role is played by {\it correlations} that are symbolically encoded in entangled states.
For example, consider the compound system $S=(S_1, S_2)$ in the state $\vert \Psi \rangle$ given by (\ref{BE1}).
Consider the projection-type observables $A_j$ on $S_j$ represented by Hermitian operators $\hat A_j$ with eigen-vectors $\vert 0\rangle,
\vert 1\rangle$ (in quibit spaces ${\cal H}_j).$  Measurement of say $A_1$ with the output $A_1=\alpha$ induces the state projection onto the vector  $\vert \alpha\alpha\rangle.$ Hence, measurement of $A_2$ will automatically produce the output $A_2=\alpha.$ 
Thus, the state  $\vert \Psi \rangle$ encodes the exact correlations for these two observables.  
In the same way, the state 
\begin{equation}
\label{BE1p}
\vert \Psi \rangle  =   (\vert 10\rangle+ \vert 01\rangle)/\sqrt{2}
\end{equation}
encodes correlations
$A_1=\alpha, A_2= 1-\alpha (\mod \; 2).$ So, {\it an entangled state provides the symbolic representation of correlations between states
of the subsystems of a compound biosystem.} 

Theory of open quantum systems operates with mixed states described by density operators. And before to turn to modeling of biological functions for compound systems, we define entanglement for mixed states. Consider the case of tensor product of two Hilbert spaces, i.e., the system $S$ is compound of two subsystems $S_1$ and $S_2.$  A mixed state of $S$ given by $\hat \rho$ is called separable if it can be   represented   as  a convex combination of product states $\hat \rho= \sum_k p_k \hat \rho_{1k} \otimes
\hat \rho_{2k},$ where  $\hat \rho{ik}, i=1,2, $ are the density operator of the subsystem $S_i$ of $S.$ Non-separable states are called entangled. They symbolically represent  correlations between subsystems. 

Quantum dynamics describes the evolution of these correlations. In the framework of  open system dynamics, a biological function approaches the steady state via the process of decoherence. As was discussed in section \ref{COG}, this dynamics resolves uncertainty 
that was initially present in the state of a biosystem; at the same time, it also washes out the correlations: the steady state 
which is diagonal  in the basis $\{\vert \alpha_1 ... \alpha_n \rangle  \}$ is separable (disentagled). However, in the  process of the state-evolution correlations between subsystems (entanglement) play the crucial role. Their presence leads to transformations of the state of the compound system $S$ 
via ``local transformations'' of the states of its subsystems. Such correlated dynamics of the global information state reflects 
{\it consistency of the transformations of the states of subsystems.} 

Since the quantum-like approach is based on the quantum information representation of systems' states, we can forget about the physical space location of biosystems and work in the information space given by  complex Hilbert space ${\cal H}.$ In this space, we can introduce the notion of locality based  on the fixed tensor product decomposition (\ref{BE0}). Operations in its components 
${\cal H}_j$ we can call  local (in information space).  But, they induce ``informationally nonlocal'' evolution of the state 
of the compound system. 

\subsection{Entanglement of genes' epimutations}
\label{ENT1}

Now, we come back to the model presented in section \ref{EPGE} and consider the information  state  of  cell's epigenome  expressing potential  epimutations of  the chromatin-marking  type.  Let cell's  genome consists of $m$ genes $g_1,...,g_m.$ For each  gene $g,$ consider all  its  possible  epimutations and   enumerate    them: $j_g = 1,..., k_g.$ The state  of all potential epimutations in the gene $g$ is represented as superposition 
\begin{equation}
\label{BEX}
\vert \psi_g\rangle= \sum_{j_g} c_{j_g} \vert j_g\rangle.
\end{equation}

In the ideal situation - epimutations of the genes are independent - the state  of  cell's  epigenome  is  mathematically described by the tensor product of the states $\vert \psi_g\rangle$:
\begin{equation}
\label{BEX1}
\vert \psi_{\rm{epi}}\rangle= \vert \psi_{g_1}\rangle \otimes ... \otimes \vert \psi_{g_m}\rangle.
\end{equation}
However,  in  a living  biosystem, the most  of  the  genes and proteins  are correlated  forming a big network system. Therefore, one epimutation   affects   other   genes. In the quantum information framework, this situation is described by entangled states:
\begin{equation}
\label{BEX2}
\vert \psi_{\rm{epi}}\rangle = \sum_{j_{g_1} ... j_{g_m}} C_{j_{g_1} ... j_{g_m}}\vert j_{g_1} ... j_{g_m}\rangle.
\end{equation}
This  form  of   representation  of potential  epimutations  in  the genome of a cell  implies  that  epimutation in one gene is consistent with epimutations in other genes. If the state is entangled (not factorized), then by acting, i.e.,  through  change  in the  environment,  to one  gene,  say $g_1,$  and  inducing some  epimutation  in  it,  the cell ``can induce'' consistent epimutations in other genes.

Linearity of the quantum information representation of the biophysical processes in a cell induces the linear state dynamics.  
This makes the  epigenetic evolution very rapid; the off-diagonal elements of the density matrix decrease exponentially quickly.
Thus, our quantum-like model justifies the high speed of the  epigenetic evolution. If it were based solely on the biopysical representation with nonlinear state dynamics, it would be essentially slower.

Modeling based on theory of open systems leads to reconsideration of interrelation between the Darwinian with Lamarckian viewpoint 
on evolution. Here we concentrated on epimutations, but in the same way we can model mutations \cite{QLX}.

\subsection{Psychological functions}
\label{ENT2}
  
Now, we turn to the model presented in section \ref{ECH}. A neural network is modeled as a compound quantum system; its state is presented in tensor product of single-neuron state spaces.  Brain's  functions perform self-measurements modeled within theory of open quantum systems. (There is no need to consider state's collapse.) State's dynamics of some brain's function (psychological function)  $F$ is described by the quantum master equation. Its steady states represent 
classical statistical mixtures of possible outputs of $F$ (decisions). Thus through interaction with electrochemical environment, 
$F$ (considered as an open system) resolves uncertainty that was originally encoded in entangled state representing uncertainties  in action potentials of neurons and correlations between them.  

Entanglement plays the crucial role in generating consistency in neurons' dynamics. As in section \ref{ENT3}, suppose that the quantum information representation is based on 0-1 code. Consider a network of $n$ neurons interacting with the surrounding electrochemical 
environment ${\cal E},$ including signaling from other neural networks. The information state is given by (\ref{BE}). 
Entanglement encodes correlations between firing of individual neurons. 
For example,  the state (\ref{BE1}) is associated with two neurons firing synchronically and the state (\ref{BE1p})   with two neurons firing asynchronically. 

Outputs of the psychological function $F$ based biophysically on  a neural network are resulted from consistent state dynamics of individual neurons belonging to this network. As was already emphasized, state's evolution towards a steady state is very rapid, as a consequence of linearity of the open system  dynamics; the off-diagonal elements of the density matrix decrease exponentially quickly.  

\section{Concluding remarks}

Since 1990th \cite{KHC1}, quantum-like modeling outside of physics, especially modeling of cognition and decision making, 
flowered worldwide. {\it Quantum information theory} (coupled to measurement and open quantum systems 
theories) is fertile ground for quantum-like flowers. The basic hypothesis presented in this paper is that functioning of  biosystems is based  on the quantum information representation of their states. This representation is the output of the biological evolution. The latter is considered as the evolution in the information space. So, biosystems react not only to material or energy constraints imposed by the environment, but also to the information constraints. In this paper, biological functions are considered as open information systems interacting with information environment. 

The quantum-like representation of information provides the possibility to process superpositions. This way of information processing is advantageous as saving computational resources: a biological function $F$ need not to resolve uncertainties encoded in superpositions and to calculate JPDs
of all compatible variables involved in the performance of $F.$ 

Another advantageous feature of quantum-like information processing 
is its linearity. Transition from nonlinear dynamics of electrochemical states to linear 
quantum-like dynamics tremendously speeds up  state-processing (for gene-expression, epimutations, and generally 
decision making). In this framework, decision makers are genes, proteins, cells, brains, ecological systems. 

Biological functions developed the ability to perfrom self-measurements, to generate outputs of their functioning. We model this ability in the framework of open quantum systems, as decision making through decoherence. We emphasize that this model is free from the ambiguous notion of collapse of the wave function.

 Correlations inside a biological function as well as between different biological functions and environment are represented linearly by entangled quantum states. 

We hope that this paper would be useful for biologists (especially working on mathematical modeling) as an introduction to the quantum-like approach to model functioning of biosystems. We also hope that it can attract attention of experts in quantum information theory to 
the possibility to use its formalism and methodology in biological studies.

\end{document}